\documentclass[twocolumn,showpacs,preprintnumbers,amsmath,amssymb]{revtex4}

\usepackage{graphicx}
\usepackage{dcolumn}
\usepackage{bm}
\usepackage{amsmath}

\begin{document}

\preprint{APS/123-QED}

\title{Microengineering laser plasma interactions at relativistic intensities}

\author{S. Jiang$^{1}$, L. L. Ji$^{1}$, H.  Audesirk$^{2}$, K. M. George$^{1}$, J. Snyder$^{1}$,  A. Krygier$^{1}$, N. S. Lewis$^{2}$, D. W. Schumacher$^{1}$, A. Pukhov$^{3}$, R. R. Freeman$^{1}$, K. U. Akli$^{1}$}

 \affiliation{$^{1}$ Department of Physics, The Ohio State University,Columbus, Ohio 43210, USA \\ $^{2}$ Division of Chemistry and Chemical Engineering, 127-72 Noyes Laboratory, California Institute of Technology, Pasadena, CA 91125. \\ $^{3}$Heinrich-Heine University of Dusseldorf, 40225, Dusseldorf, Germany}

\date{\today}

\begin{abstract}
We report on the first successful proof-of-principle experiment to manipulate laser-matter interactions on the microscales using highly ordered Si microwire arrays. The interaction of a high contrast short pulse laser with a flat target via periodic Si microwires yields a substantial enhancement in both total and cut-off energies of the produced electron beam. The self generated electric and magnetic fields behave as an electromagnetic lens that confines and guides electrons between the microwires as they acquire relativistic energies via direct laser acceleration (DLA).


\end{abstract}
\maketitle

Laser matter interactions at relativistic intensities have exhibited many interesting physical processes.  These include the acceleration of electrons~\cite{Faure,Geddes,Wharton,Malka}, protons, and heavy ions~\cite{Snavely, Hegelich, Habara}, creation of electron-positron jets~\cite{Cowan, Gahn,Chen}, and attosecond pulse generation~\cite{Dromey, Thaury}. The investigation of ultra-short pulse lasers interacting with initially solid-density matter has been mainly focused on flat targets with little or no control over the interaction. Recently the focus has shifted toward using advanced targets with the aim of increasing laser beam absorption and subsequent energy partition among various plasma species. Structured interfaces including nanoparticles~\cite{Rajeev}, snowflakes~\cite{Zigler}, and nanospheres~\cite{Margarone}  have been reported to enhance laser absorption and proton acceleration. Trapping of femtosecond laser pulses of relativistic intensity deep within ordered nanowires resulted in volumetric heating of dense matter into a new ultra-hot plasma regime~\cite{Purvis}. The potential of prescribed geometrical structures on the front of a target to greatly enhance the yield of high-energy electrons while simultaneously confining the emission to narrow angular cones has been proposed~\cite{Jiang}.

Microengineering laser plasma interactions, at intensities above the material damage threshold, has not been extensively explored. The main reason is that the amplified short pulses are inherently preceded by nanosecond scale pedestals~\cite{Ivanov}. This departure from an ideal pulse can substantially modify or destroy any guiding features before the arrival of the intense portion of the pulse. The introduction of plasma mirrors for beam cleaning improves the contrast between the main laser pulse and the unwanted pre-pulse intensity, but do so at a steep cost in overall laser energy throughput and mode quality. 

Laser pulse cleaning techniques are now being employed to significantly minimize unwanted pre-pulse and pedestals. For example, Ti-Saph based short-pulse high-intensity lasers routinely use cross-polarized wave generation (XPW) technique to achieve a contrast of at least $10^{10}$ on the nanosecond time scale~\cite{Jullien}. The manufacturing of advanced micro and nanostructures has been the domain of specialized scientific disciplines such as nanoelectronics~\cite{Min}, microfluidics~\cite{Unger}, and photovoltaic~\cite{Jang}. Microstructures with features as small as 200 nm can now be easily manufactured by non-experts using commercially available 3D direct laser writing (DLW) instruments~\cite{Fischer}. Furthermore, 3D large-scale simulations with enough spatial and temporal resolutions to capture the details of the interaction are now possible thanks to recent advances in massively parallel computing capabilities coupled with newly developed Particle-In-Cell (PIC) algorithms. 

In this letter, we first investigate the interaction of a high contrast ultra-short pulse with highly ordered microwire arrays using the Virtual Laser Plasma Laboratory (VLPL) code in full 3D geometry~\cite{VLPL}. We identify the mechanisms of relativistic electron beam acceleration and guiding. Then, we report on the first successful proof-of-principle experiment to produce and control laser-driven electrons with nanoassembled Si microwires.

\begin{figure*}
\includegraphics[width=140mm,height=160mm]{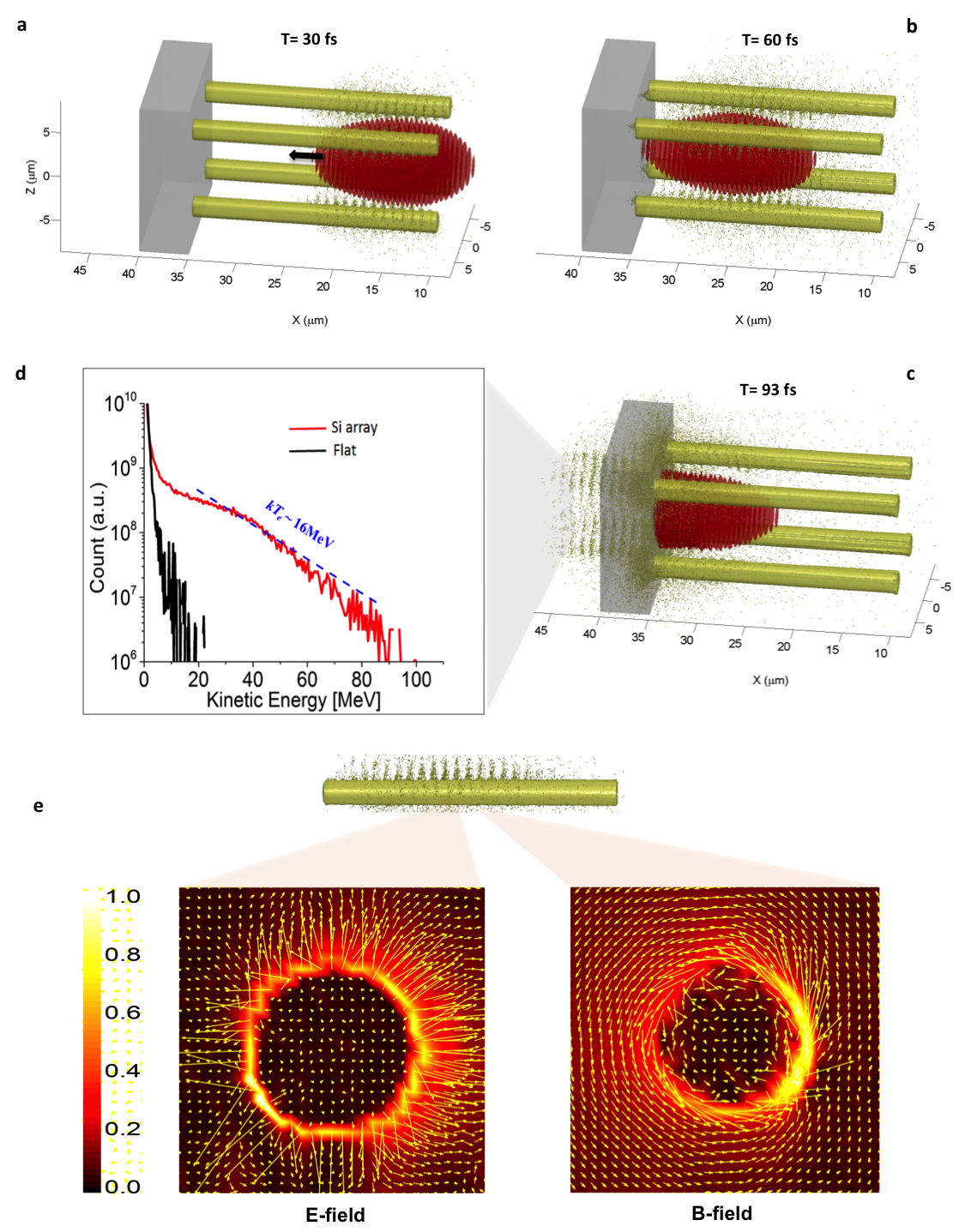}
\caption{ \textbf{$\mid$ Three-dimensional PIC simulations of an intense short-pulse laser interacting with microwire arrays}: laser pulse ($I = 10^{21} \ Wcm^{-2}$) is polarized in the y direction and propagates along the x direction from right to left. The target consists of highly aligned periodic carbon wires 20 $\mu$m long, 1.2 $\mu$m thick attached to 5.6 $\mu$m thick aluminum substrate. a, b, and c are snapshots of the interaction at t=30fs, t= 60fs, and t=93fs respectively. d, electron beam energy distributions from microwire array (red) and flat Al target (black). e, electric and magnetic fields around the wire, average overg the wire length.\label{fig:wide}}
\label{fig:Sim2} 
\end{figure*}

Our PIC investigation was carried out in a simulation box with $48 \lambda_0 \times 20 \lambda_0 \times 20 \lambda_0$  in $x \times y \times z$ directions respectively ($\lambda_0 = 0.8 \mu m$ is the laser wavelength). A laser pulse polarized in the $y$ direction enters the simulation box from the left boundary along the $x$ direction. The laser field amplitude has a profile of $a_y = a_0e^{-(r/\sigma_0)^2}\sin{(\tfrac{\pi t}{2 \tau_0})}$ where $\omega_0$ is the laser frequency, $a_0 = eE_L/m_e \omega_0 c$ is the dimensionless laser electric field amplitude. Here $e$, $m_e$ are fundamental charge and electron mass, $E_L$ is the laser electric field, $c$ is the speed of light in vacuum, respectively. The laser amplitude, pulse duration, and spot size are $a_0 = 21$, $\tau_0 \approx 40 fs$ and $\sigma_0 = 4 \lambda_0$ respectively. Periodic carbon microwires with a length of $25 \lambda_0$, diameter of $1.5 \lambda_0$, and spatial spacing of $7 \lambda_0$ are placed $10 \lambda_0$ from the left boundary. The electron density of the wires is $n_e = 300 n_c$ and they are attached to an Aluminum foil of $n_{Al} = 25 n_c$ density ($n_{c}=m_{e}\omega_{0}^{2}/4\pi e^{2}$ is the critical density). The entire target is cold and pre-ionized.

Snapshots of the interaction and electron beam energy distributions from simulations are shown in Figure~\ref{fig:Sim2}.  As the pulse enters the microwire array (Fig.~\ref{fig:Sim2}a), electrons are pulled out of the wires by the laser field. These electron bunches are periodic and they are separated by one laser wavelength on each wire. The electron bunches on two opposite wires are separated by half laser wavelength reflecting the oscillatory nature of the driving laser field. The laser pulse has a phase velocity approximately equal to the speed of light as it propagates between the wires (Fig.~\ref{fig:Sim2}b). Consequently,  electrons pulled from the surface of the wires are injected into the laser pulse and accelerated via direct laser acceleration mechanism (DLA)~\cite{DLA}. Finally, when the laser beam reaches the flat interface, electrons originating from the wires have acquired significant kinetic energy. They propagate in the forward direction and escape the target as indicated by the green periodic bunches at the rear side of the target (Fig.~\ref{fig:Sim2}c). It is worth noting that lower energy electrons are produced as well when the pulse irradiates the flat surface holding the wires. The most energetic electrons are the ones that originated from the wires and experienced acceleration via DLA. Fig.~\ref{fig:Sim2}d shows the electron energy distribution for the microwire array target. As a baseline comparison, we have carried out simulations of a flat interface without the wires using the same laser and simulation parameters. It is clear from these results that the performance of microengineered targets is superior to that of a flat target in producing and accelerating electrons. In the microwire array target, electrons with energies as high as 90 MeV are produced compared to 20 MeV maximum electron energies in flat targets. An exponential fit to the electron energy distribution from the microwire array target yields $kT_{e}$=16  MeV, much higher than the ponderomotive scaling at the same intensity ($kT_{e}$=7  MeV)~\cite{Wilks}. The total number of relativistic electrons with energies higher than 1 MeV is enhanced by a factor of 25 with the structured interface compared to flat targets. We also observe that the accelerated electrons travel forward in the vicinity of the wires. This suggest the presence of a guiding mechanism induced by laser plasma interactions. 

We examined the electric and magnetic fields in the neighborhood of the wires. Fig.~\ref{fig:Sim2}e (left) shows a vector plot of the electric field surrounding one representative wire. The electric field is radially oriented and points away from the wire. This is consistent with fields induced by a distribution of positive charge. In our case, this field is induced by charge separation as the electrons are pulled from the wires by the laser, leaving the ions behind. Fig.~\ref{fig:Sim2}e (right) shows the magnetic field in the vicinity of the wire. The orientation of the magnetic field and field line configuration is consistent with the magnetic field of current carrying wire. The magnetic field is induced by the return currents in the wire as the plasma responds to electric current unbalance produced by the forward going super-thermal electrons. These plasma-produced electric and magnetic fields provide a guiding mechanism for electrons that are accelerated by DLA. The electric field induced by charge separation tends to attract electrons toward the wires. The magnetic field tends to push them away, toward the laser axis. Electrons with velocities such that the transverse electric and magnetic forces cancel one another are guided in the forward direction. The simulation results suggest that these advanced targets can be used as microphotonic devices to manipulate light matter interaction on small scales and subsequently control the production of secondary particle beams.

\begin{figure}[t]
\begin{center}
 \includegraphics[angle=0, height=100 mm]{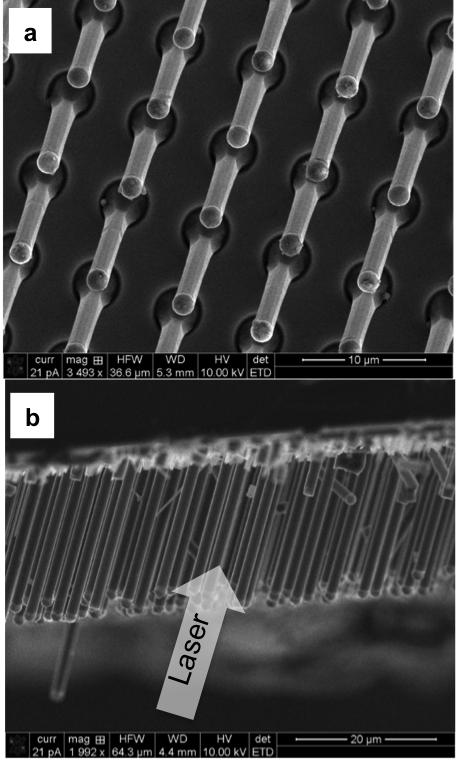}
\caption{\textbf{$\mid$ A scanning electron microscope (SEM) images of microphotonics targets:} a, top view showing wire spatial distribution. b, side view showing the orientation of the wires with respect to the 450 $\mu$m thick flat Si substrate. Laser is incident parallel to the wires (white arrow)}
\label{fig:T}
\end{center}
\end{figure}

\begin{figure*}
\includegraphics[width=170mm,height=70mm]{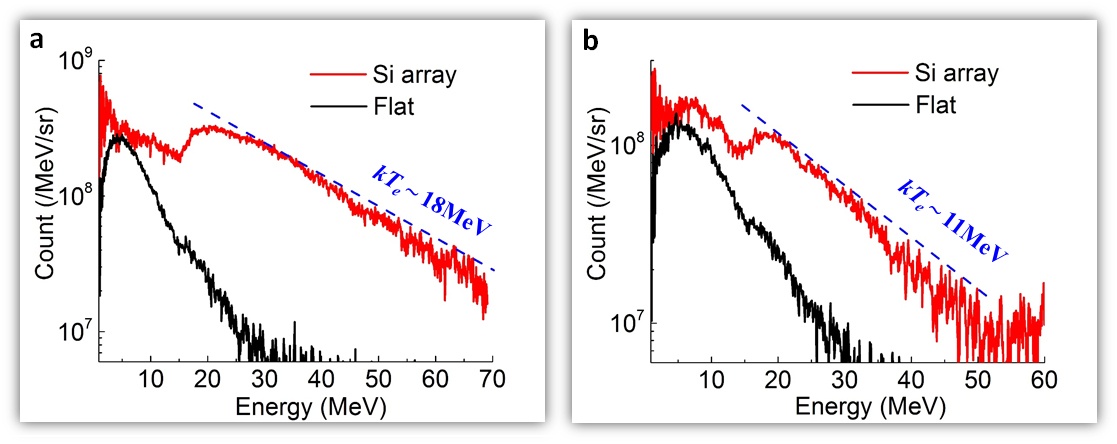}
\caption{ \textbf{$\mid$ Experimental results}: Escaping electrons energy distributions for 4 laser shots. Flat target spectra (black), Si arrays spectra (red). a) Si array target-1 and baseline flat substrate. b) Si array target-2 and baseline flat substrate.
\label{fig:wide}}
\label{fig:Ex2} 
\end{figure*}

A proof-of-principle experiment was conducted on the Scarlet laser facility at The Ohio State University~\cite{Poole}. To manipulate light-matter interactions, we used Si microwire arrays~\cite{Warren}. Si$\{211\}$ substrates were used to grow the periodic inclined Si wires. The surface of the substrate is first oxidized to form a layer of SiO$_{2}$ a few hundred nanometers thickness. Then a thin layer of photoresist is applied using spin coating. The position and diameter of the microwires are prescribed by creating circular holes on the photoresist layer via photolithography. Using a buffered HF etching, the uncovered SiO$_{2}$ under the holes is removed to expose the pure Si. The holes are then filled up with a few hundred nanometers of Cu via thermal evaporation onto the photoresist, followed by liftoff to dispose of the photoresist layer. The substrate is then annealed and Si wires are grown through vapor-liquid-solid (VLS) growth method, while other portion of the surface is still protected by the SiO$_{2}$ layer. 
Silicon wires with 1.5 $\mu$m diameter, 15-25 $\mu$m  length, and 7 $\mu$m spacing were grown on a 450 $\mu$m thick flat Si substrates  (Fig.~\ref{fig:T}a). 

The laser delivered 4-5J of energy on target with main pulse to amplified spontaneous emission (ASE) intensity contrast better than $10^{9}$. The 40 fs duration laser pulse was focused with an F/2.2 off-axis parabola to a 3 $\mu$m full width at half maximum focal spot, reaching a peak intensity $\sim 1\times10^{21}$ Wcm$^{-2}$.  To prevent laser back-reflections from damaging the front-end optics, the wires were grown at 22.5 degrees with respect to the flat substrate normal (Fig.~\ref{fig:T}b). The laser propagation direction was parallel to the wires and electrons escaping the rear side of the target were collect with a  magnetic spectrometer coupled to imaging plate detectors. The magnetic field in the center of the gap of our spectrometer is about 0.6 T  and the instrument collected electrons at 30 degrees from the laser axis. 

The experimental results are summarized in Fig.~\ref{fig:Ex2}. Electron beam energy distributions from Si microwire arrays are shown in red and the distributions from flat targets are shown in black. The electron beam energy distributions recorded with flat interfaces are similar. In both Fig.~\ref{fig:Ex2}a and Fig.~\ref{fig:Ex2}b, the cut-off energy for the electron beam is around 30 MeV. Our PIC simulations predicted a maximum energy of 20 MeV. A significant enhancement in total number of electrons and their mean energy is obtained with Si microwire arrays. The structured target in Fig.~\ref{fig:Ex2}a yielded cut-off energy of 70 MeV while the structured target in Fig.~\ref{fig:Ex2}b yielded 60 MeV. For both structured targets, two electron populations characterize the spectra: a low energy population in the range of 0.5-20 MeV and a high electron energy population that peaks around 25 MeV and extends to 60-70 MeV range. This suggest, as seen in the simulations, that the spectrum is a combination of electrons from the bulk of the target and electrons from the wires that were accelerated by DLA. An exponential fit to the data gives $kT_{e}$=18 MeV, and $kT_{e}$=11  MeV for data in  Fig.~\ref{fig:Ex2}a and Fig.~\ref{fig:Ex2}b respectively. This is consistent with the predictions of our 3D PIC simulations ($kT_{e}$=16  MeV).

In conclusion, we report on first successful proof-of-principle demonstration of microengineering laser plasma interactions. Specifically, shaping laser-driven electron beam characteristics. By manipulating light-matter interactions on the micro and nanoscales, various processes such as ion acceleration and electron-positron pair production can be controlled and optimized. This work brings nano and microscience to high field physics and will open new paths towards plasma microphotonics with ultraintense tabletop lasers.

\begin{acknowledgments}
KUA and SJ would like to thank C. Willis, P. Poole, R. Daskalova, and E. Chowdhury for their assistance during the experimental run. This work is supported by the AFOSR Basic Research Initiative (BRI) under contract FA9550-14-1-0085 and allocations of computing time from the Ohio Supercomputing Center. A. Pu is supported by DFG Trnsregio TR18 (Germany).  NSL and HA acknowledge the National Science Foundation, grant CHE-1214152, for support.
\end{acknowledgments}


\end{document}